\documentclass[amssymb,superscriptaddress,showpacs,twocolumn,prd]{revtex4}
\usepackage{graphicx}

\newcommand{\beq}[1]{\begin{eqnarray}\label{#1}}
\newcommand{\eeq}{\end{eqnarray}}

\begin{document}
 \pagestyle{plain}

 \title{Physics Related with Co-moving Coordinate System}

 \author{Ding-fang Zeng}
 \email{dfzeng@itp.ac.cn}
 \affiliation{Institute of Theoretical
 Physics, Chinese Academy of Science.}
 \begin{abstract}
 We derive the metric of an expanding universe with zero
 accelerations by pure kinematic method. By doing so we expatiate
 physics related with co-moving coordinate system in details. The
 most important discovery or our study is, in an expanding
 universe with zero accelerations, the red-shift of photons from
 distance galaxies is determined by the co-moving coordinate of
 the source galaxy instead of the scale factor's time dependence.
 Our discovery is consistent with the current observed
 super-novaes's luminosity-distance v.s. red-shift relations.
 \end{abstract}


 \maketitle

 \section{Introduction}

 The luminosity-distance v.s. red-shift relation of type Ia super-novae
 is the most direct evidence of dark energy's existence
 see \cite{Riess98,Perlmutter98,Knop03,Tonry03,Riess04} for experimental
 literatures and \cite{Quintessence2,Phantom,Phantom2,backReaction}
 for theoretical explanations. The
 matter distribution power spectrum observed by
 SDSS \cite{Tegmark04} and cosmic microwave
 background anisotropy observed by WMAP \cite{WMAP03} is indirect evidence of
 dark energy's existence.
 We will provide a new explanation for super-novae's
 luminosity-distance v.s. red-shift relation without assuming that
 the universe is accelerate-ly expanding. Our explanation will avoid
 almost all the main problems of standard cosmology before inflation
 is introduced into physics.

 Our explanation resorts to new interpretation of physics
 related with co-moving coordinate system. As the first step,
 We would like to ask, if one
 focuses on a given direction of a non-perturbed
 universe, will he see an infinitely long, uniform and expanding
 galaxy line? If no, why? We consider only flat universe.

\begin{figure}[h]
 \vspace{-5mm}
 \begin{minipage}[]{0.9\textwidth}\includegraphics[]{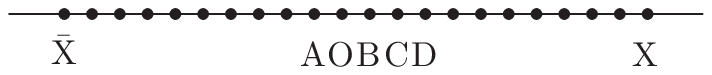}\end{minipage}
 \vspace{3mm}
 \caption{an infinitely long uniform galaxy line}
 \label{LorentzContraction}
 \end{figure}

 If yes, suppose this man/woman were put on galaxy
 $O$ and were asked to measure the recession velocity of galaxy
 $B$ and $C$, see FIG.\ref{LorentzContraction},
 what result will he/she get? $(v,2v)$ or $(v,\frac{2v}{1+v\cdot
 v})$? $v$ is the relative recession velocity between two nearest
 galaxies.
 We insist the second answer, i.e., we insist that (i)
 cosmological principle is a local statement;
 (ii) the definition of simultaneity can only be relativistic.

 If the one dimensional system in Figure.\ref{LorentzContraction}
 is uniformly expanding, the metric is
 \beq{}
 ds^2=-dt^2+a^2(t)dx_{co}^2,\label{metricD1sCDMstyle}
 \eeq
 when generalizing into (1+3)D space-time, we have
 \beq{}
 ds^2=-dt^2+a^2(t)(dr_{co}^2+r^2_{co}[d\theta^2+\sin^2\theta d\phi^2])
 \label{metricD3sCDMstyle}
 \eeq
 Some standard cosmologists claim that, the generalization
 of $(1+1)D\Rightarrow(1+3)D$ is
 ir-rationale. Because (1+1)D gravitation
 theory is topological, its $G_{\mu\nu}==0$, so no dynamical
 equations can be used to determine $a(t)$.
 However if we know the (1+1)D $a(t)$ somehow, the
 generalization to (1+3)D is rationale, because if we
 are considering a non-perturbed universe
 and if we are focusing on only a given direction,
 we will see the galaxy line in FIG. \ref{LorentzContraction}. Since the projection
 $(1+3)D\Rightarrow(1+1)D$ involves only kinematic, it involves no
 dynamics.

 So our claiming is: if we know the (1+1)D $a(t)$ somehow,
 generalization of eq(\ref{metricD1sCDMstyle}) into
 eq(\ref{metricD3sCDMstyle}) is rationale, because the
 generalization only involves kinematic, it involves no
 dynamics. Although the (1+1)D gravitation is topological, $G_{\mu\nu}==0$,
 the (1+3)D metric obtained by generalizing a pre-given (1+1)D metric has
 non-zero $G_{\mu\nu}$. So non-trivial dynamics can appear in the
 generalized (1+3)D space-time.

 For example, if the system illustrated in FIG. \ref{LorentzContraction}
 is expanding with zero acceleration, we can derive
 its metric --(1+1)D form-- by pure kinematic method then generalize the
 results into (1+3)D space-time thus obtain the metric of an
 isotropic and homogeneous universe whose expansion has zero
 acceleration.
 The (1+1)D metric we obtained by pure kinematic
 method has identically zero $G_{\mu\nu}$, but the generalized
 (1+3)D metric has non-zero $G_{\mu\nu}$. So in the (1+3)D space-time,
 non-trivial dynamic appears. On the other hand,
 if we know the energy momentum tensor corresponding with an
 expanding universe with zero accelerations as priors,
 directly solving the (1+3)D dynamic equation will give
 us the same metric as
 that obtained by kinematic method of $(1+1)D\Rightarrow(1+3)D$.

 \section{(1+1)D Expanding Universe with Zero Accelerations}

 Take the galaxy line in FIG.\ref{LorentzContraction} as
 our experimental labs. Suppose the system is expanding
 with zero-accelerations. Considering the following series
 \beq{}
 &&\hspace{-3mm}v_B=v;\nonumber\\
 &&\hspace{-3mm}v_C=\frac{v+v}{1+v^2};\nonumber\\
 &&\hspace{-3mm}v_D=\frac{v+v_C}{1+v\cdot v_C};\nonumber\\
 &&\textrm{... ...}\nonumber\\
 &&\hspace{-3mm}v_X=\frac{v+v_{X-1}}{1+v\cdot v_{X-1}};
 \label{HubbleSeries}\\
 &&\hspace{-3mm}|AB|=2a\nonumber\\
 &&\hspace{-3mm}|OC|=2a\sqrt{1-v_B^2}\nonumber\\
 &&\hspace{-3mm}|BD|=2a\sqrt{1-v_C^2}\nonumber\\
 &&\textrm{... ...}\nonumber\\
 &&\hspace{-3mm}|X^-X^+|=2a\sqrt{1-v_X^2},
 \label{LorentzSeries}
 \eeq
 $v$, $a$ are locally measured relative recession velocity and distance between two nearest
 galaxies respectively. Since we consider only
 the non-accelerate-ly expanding universe, so
 $a=v\cdot t$. From series eqs(\ref{HubbleSeries}) and (\ref{LorentzSeries}) we get
 \beq{}
 v_X&&\hspace{-3mm}=\frac{(1+v)^X-(1-v)^X}{(1+v)^X+(1-v)^X};
 \label{recessionVelocity}\\
 |OX|&&\hspace{-3mm}\sim a\sum_{N=0}^{X}{\sqrt{1-v_N^2}}\nonumber\\
 &&\hspace{-3mm}=l\int_0^X dx\sqrt{1-v_x^2}\nonumber\\
 &&\hspace{-3mm}=\frac{4a}{\text{ln}\frac{1+v}{1-v}}\left[\text{arctg}[(\frac{1+v}{1-v})^{\frac{X}{2}}]-\frac{\pi}{4}\right],
 \label{physicalDistance}
 \eeq

 From eq(\ref{physicalDistance}) using light velocity invariance
 principle, we can write down the metric of our (1+1)D
 non-accelerate-ly expanding universe as
 \beq{}
 ds^2=-dt^2+\frac{4v^2t^2}{(e^{\sigma x}+e^{-\sigma x})^2}dx^2
 \label{metricDim1Nat}\\
 \text{where }
 \sigma=\frac{1}{2}\text{ln}\frac{1+v}{1-v}\label{sigmaDefinition}
 \eeq
 or
 \beq{}
 ds^2=-dt^2+a^2(t)dx_{co}^2,\ a(t)=v\cdot t\label{metricDim1Co}\\
 \text{where}\ x_{co}=\frac{2}{\sigma}(\text{arctg}[e^{\sigma x}]-\frac{\pi}{4}).
 \eeq

 We call the coordinate $x$ in eq(\ref{metricDim1Nat}) natural
 coordinate, while the
 coordinate $x_{co}$ in eq(\ref{metricDim1Co}) co-moving
 coordinate. Natural coordinate ranges in $(-\infty, \infty)$, but
 co-moving coordinate only ranges in
 $(-\frac{\pi}{2\sigma},\frac{\pi}{2\sigma})$. If $v\rightarrow0$,
 natural coordinate coincides with co-moving coordinate.
 The co-moving coordinate definition of Standard
 cosmology emphasizes only one point: co-moving coordinate is a
 coordinate fixed on galaxies, the co-moving coordinate of a given galaxy
 does not vary as background universe expands. By this definition,
 natural coordinate is also co-moving coordinate.

 But the difference between co-moving coordinate and
 natural co-ordinate is very important. Natural coordinate is related with
 physical coordinate through
 \beq{}
 x_{ph}=\frac{2vt}{\sigma}(\text{arctg}[e^{\sigma x}]-\frac{\pi}{4}).
 \label{xphxNatRelation}
 \eeq
 While co-moving coordinate is related with physical coordinate
 through
 \beq{}
 x_{ph}=a(t)\cdot x_{co},\ a(t)=v\cdot t.
 \label{xphxcoRelation}
 \eeq
 Although the co-moving coordinate defined in
 eq(\ref{xphxcoRelation}) is very similar to that
 of standard cosmology, it has completely different
 interpretation from that of standard cosmology.

 By standard cosmology's definition $x_{ph}=a(t)\cdot x_{co}$, if
 we have a photon emitted at $(t,x_{co})$ and
 detected at $(t_0,0)$, the red-shift of this photon
 is $(1+z)=\frac{a(t_0)}{a(t)}$, it has nothing to do with
 the co-moving coordinate of source galaxy.
 But in the co-moving coordinate definition of
 eq(\ref{xphxcoRelation}), for the same photon, the
 red-shift is
 \beq{}
 (1+z)=\sqrt{\frac{1+v_x}{1-v_x}}=e^{\sigma x}
 \label{redShiftDim1}
 \eeq
 It is completely determined by the co-moving coordinate
 of source galaxy but has nothing to do with scale factor!

 If we accept eqs(\ref{xphxNatRelation})+(\ref{redShiftDim1}),
 then even without assuming that the universe is accelerate-ly
 expanding, we can give the observed luminosity-distance v.s.
 red-shift relation of super-novaes a very beautiful explanation.
 Of course, if one would like to, he can replace the velocity $v$ in
 eqs(\ref{metricDim1Nat}), (\ref{metricDim1Co}),
 (\ref{xphxNatRelation}) and (\ref{xphxcoRelation}) with a time
 dependent function $v(t)=v+pt+\frac{1}{2}qt^2+...$, and the factor
 $a(t)=v\cdot t+\frac{1}{2}pt^2+\frac{1}{3!}qt^3+...$, thus obtain
 appropriate relations in a accelerate-ly expanding universe.
 In this case eq(\ref{redShiftDim1}) will be changed so that more
 free parameters enter, hence more precise fitting with
 experiments can be obtained.

 \section{Generalization into (1+3)D Space-time}

 Generalizing eqs(\ref{metricDim1Nat}) or (\ref{metricDim1Co})
 into (1+3)D space-time, we obtain
 \beq{}
 r_{ph}&&\hspace{-3mm}=\frac{2vt}{\sigma}(\text{arctg}e^{\sigma
 r}-\frac{\pi}{4})\label{rph_rna_relation}\\
 ds^2&&\hspace{-3mm}=-dt^2+\frac{v^2t^2}{\cosh^2{\sigma r}}(dr^2+r^2[d\theta^2+\sin^2\theta
 d\phi^2])
 \label{metricDim3Nat}\\
 r_{ph}&&\hspace{-3mm}=a(t)\cdot r_{co},\ a(t)=vt,0\leq r_{co}\leq\frac{\pi}{2\sigma}\\
 ds^2&&\hspace{-3mm}=-dt^2+a^2(t)(dr_{co}^2+r_{co}^2[d\theta^2+\sin^2\theta d\phi^2])
 \label{metricDim3co}
 \eeq
 The parameter $v$ now should be understood as the
 average recession velocity between two nearest galaxies.
 If we take a limit $v\rightarrow0$, co-moving coordinate reduce
 to natural coordinate, eq(\ref{metricDim3co}) becomes
 (\ref{metricDim3Nat}).
 Note, we consider only galaxies which are performing Hubble
 recessions relative to each other. We do not consider galaxies
 bounded in the galaxy clusters.

 Just the same as (1+1)D case,
 eq(\ref{metricDim3co}) is very similar to standard cosmology's
 FRW metric, but the two has completely different physical
 interpretations.
 E.g. if we have a photon emitted at
 $(t,r_{co},\theta,\phi)$ and detected at $(t_0,0,\theta,\phi)$.
 By standard cosmology, the red-shift of this photon is
 $(1+z)=\frac{a(t_0)}{a(t)}$; but by our explanation of
 eq(\ref{metricDim3co}), the red-shift is
 \beq{}
 (1+z)=\sqrt{\frac{1+v_r}{1-v_r}}=e^{\sigma r}
 \label{redShiftDim3}
 \eeq

 Although our starting point, the (1+1)D metric
 eqs(\ref{metricDim1Nat})
 and (\ref{metricDim1Co}) are topological theory, it has no
 dynamics. When we generalize them into (1+3)D
 case, eqs(\ref{metricDim3Nat}) or (\ref{metricDim3co}),
 non-trivial dynamics appears. By Einstein equation, we can
 calculate the energy momentum tensor corresponding with them.
 The results are respectively
 \beq{}
 8\pi GT_{\mu\nu,na}&&\hspace{-3mm}=-G_{\mu\nu,na}=-\text{diag}
 \nonumber\\
 &&\hspace{-15mm}
 \left.\{
 \frac{-(6v^2r+5\sigma^2r-\sigma^2r\cosh[2\sigma r]+4\sigma\sinh[2\sigma r])}
  {2v^2rt^2}
 \right.\nonumber\\
 &&\hspace{-12mm}
 \left.
 ,(\sigma^2+v^2)\text{sech}^2[\sigma r]+\sigma(-\sigma+\frac{2\tanh[\sigma r]}{r})
 \right.\nonumber\\
 &&\hspace{-8mm}
 \left.
 ,r^2(\sigma^2+v^2)\text{sech}^2[\sigma r]+\sigma r\tanh[\sigma r],
 \right.\nonumber\\
 &&\hspace{-10mm}
 \left.
 \sin^2\theta\left[r^2(\sigma^2+v^2)\text{sech}^2[\sigma r]+\sigma r\tanh[\sigma
 r]\right]
 \right\}
 \label{EMTdim3Nat}\\
  8\pi GT_{\mu\nu,co}&&\hspace{-3mm}=-G_{\mu\nu,co}\nonumber\\
 &&\hspace{-3mm}=-\text{diag}
 \{ -\frac{3}{t^2},v^2,v^2r_{co}^2,v^2r_{co}^2\sin^2\theta \}\label{EMTdim3co}
 \eeq

 If we know the energy momentum tensor describing the cosmological fluid is
 $T^\mu_\nu=\text{diag}\{\rho,\frac{1}{3}\rho,\frac{1}{3}\rho,\frac{1}{3}\rho\}$
 in the co-moving coordinate, i.e. $T_{\mu\nu}$ expressed in
 eq(\ref{EMTdim3co}) times $g^{\mu\nu}$, then starting from a general ansaltz
 $ds^2=-dt^2+a(t)(dr_{co}^2+r_{co}^2d\Omega^2)$, using Einstein equation, we can also
 derive out the function form of $a(t)=v\cdot t$.
 This is just the routine of standard cosmology. But
 this routine slides over all physics related with the definition
 of co-moving coordinate system, see eq(\ref{redShiftDim3}) and
 the related remarks.

 From eq(\ref{EMTdim3co}), we can see that in the co-moving
 coordinate system, for a non-accelerately expanding universe, its cosmological
 fluid has pressure $p=-\frac{1}{3}\rho$. Note it is $T^\mu_\nu$, not
 $T_{\mu\nu}$,
 that is directly related with energy density and pressure,
 $T^\mu_\nu=\rho u^\mu u_\nu+p(u^\mu u_\nu+\delta^\mu_\nu)$.
 We have two reasons to accept this negative
 pressure.

 The first reason is, we can think it originates from dark
 energy, e.g.,
 $T^\mu_\nu$ = $(\rho,\frac{1}{3}\rho,\frac{1}{3}\rho,\frac{1}{3}\rho)$ =
 $(\frac{2}{3}\rho,0,0,0)$ +
 $(\frac{1}{3}\rho,\frac{1}{3}\rho,\frac{1}{3}\rho,\frac{1}{3}\rho)$.
 Note, in standard cosmology, to prevent the expansion of universe from
 decelerating, dark energy must be included in the total energy
 menu of the universe.

 The second reason is, even in a universe containing only matters,
 negative pressure can appear as a result of gravitations.
 Imagine an infinitely long uniform
 galaxy line, if one of the composite galaxies is less weighted than others,
 then all galaxies on the left hand side of this less weighted galaxy will
 collapse and move to the left, while all galaxies on the right hand side of
 this less weighted galaxy will collapse and move to the right.
 So, the less weighted galaxy receives gravitations which
 have intentions to split it into two parts. This intention can
 be understood as the origin of negative pressure.
 A less weighted galaxy is just an auxiliary object to
 illustrate the effects negative pressure. When all
 galaxies are equal weighted, negative pressure also exists as a result
 of gravitations.

 If we accept the first reason, i.e., negative pressure originates
 from dark energy, then we will have to accept that our metric
 eqs(\ref{metricDim3Nat}) and (\ref{metricDim3co}) only describe
 our universe in a very short period of time. Because $\rho_{m}\propto
 a^{-3}(t)$, while $\rho_{de}\propto a^{-3(1+w)}(t)$. If
 $\rho_{m}\sim\rho_{de}$ today, then in the far past, $\rho_{de}$
 must be much less than $\rho_{m}$, so negative pressure provided
 by dark energy will not be able to prevent the universe from
 decelerate-ly expanding.
 If we accept the second reason, i.e. dark energy originates from
 gravitations between different parts of the universe, then
 our metric eqs(\ref{metricDim3Nat}) and
 (\ref{metricDim3co}) can be used to describe the universe
 in any eras when the gravitation is the main interaction
 between different parts of the universe.

 Of course, at very early times, galaxies do not exist, so the
 parameter $v$ cannot be understood as the average recession
 velocity between two nearest galaxies, but according to the hierarchical
 clustering scenario, before galaxies appear, stars exist, before
 stars appear, nucleon exists, before nucleon appears, electron
 and protons exists, ... . So, as long as we
 accept that inter-gravitations among different parts of the universe
 produce negative pressures, while the average velocity of relative recession
 between two nearest composite object is $v$,
 then eqs(\ref{metricDim3Nat}) and
 (\ref{metricDim3co}) can be used to describe our universe at
 times
 as early as big-bang nucleon synthesis, even primordial
 singular point. While God, need only calculate and assign
 value to one parameter $v$, so that when the universe is 137Gyr
 old, human appears.

 \section{Observations of Super-novae}
 From theoretical aspects, the basis of eqs(\ref{metricDim3Nat}) and
 (\ref{metricDim3co}) is very simple and concrete, (i)
 cosmological principle is a local statement; (ii) the definition
 of simultaneity can only be relativistic. However, will experimental observations support
 it? The luminosity-distance v.s. red-shift relation of super-novaes
 is the most direct evidence that the universe is accelerate-ly
 expanding. But this statement is based on ignoring of
 physics related with co-moving coordinate
 system discovered in this paper. We will show that when
 considering physics related with co-moving coordinate system,
 the observational result can be explained even without assuming
 that our universe is accelerate-ly expanding.

 If we consider physics related with co-moving coordinate system,
 the red-shift of photons coming from distant galaxies will be
 changed remarkably comparing with standard cosmology. When the
 universe is assumed expanding with zero acceleration, photons
 emitted from a super-novae at position $(t,r,\theta,\phi)$ have
 red-shift
 \beq{}
 (1+z)=\sqrt{\frac{1+v_r}{1-v_r}}=e^{\sigma r}.
 \label{redShift2Dim3}
 \eeq
 Considering Lorentz dilating, the photons emitted in period $\delta
 t_1$ can only reach us in period $\delta
 t_1e^{\sigma r}$. So we get the luminosity-distance v.s.
 red-shift relation as
 \beq{}
 d_l=(1+z)\cdot\frac{2v\cdot H_0^{-1}}{\sigma}[\text{arctg}(1+z)-\frac{\pi}{4}]
 \label{dlzRelationForumlae}
 \eeq
 Please refer to \cite{SWeinberg}, section 14.4, eqs(14.4.11-14)
 for detailed derivation of eq(\ref{dlzRelationForumlae}).

 \begin{figure}[h]
 \hspace{7mm}\includegraphics[width=80mm,height=60mm]{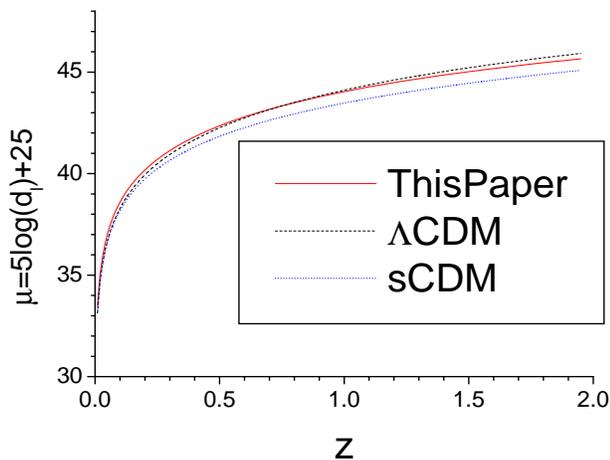}
 \caption{
 The luminosity distance v.s. red-shift relation of super-novaes.
 Red(solid) line is the prediction of this paper;
 Black(dot) line is the prediction of $\Lambda$CDM cosmology, in
 which $\Omega_{m0}=0.27$, $\Omega_{\Lambda}=0.73$, $H_0=71\textrm{km/(s}\cdot\textrm{Mpc)}$;
 Blue(dash) line is the prediction of standard CDM cosmology, in
 which $\Omega_{m0}=1.0$,
 $H_0=71\textrm{km/(s}\cdot\textrm{Mpc)}$.
 }
 \label{Observations}
 \end{figure}

 From FIG. \ref{Observations}, we can see that when considering
 physics related with co-moving coordinate system, even without
 assuming that our universe is accelerate-ly expanding,
 theoretical predictions are very close to predictions of
 $\Lambda$CDM cosmology.
 From best fitting observational results
 of \cite{Riess04}, we get $v=0.79/3000$,
 $H_0=60\text{km/(s}\cdot\text{Mpc)}$, $\chi^2=303$ (186\text{Golden+Silver
 sample}) or $v=0.899/3000$,
 $H_0=$$60\text{km/(s}\cdot\text{Mpc)}$, $\chi^2=237$ (157\text{Golden
 sample}).

 Only judging from numerical fitting qualities, our prediction
 eq(\ref{dlzRelationForumlae}) may
 be not as good as standard cosmology. But our theoretical
 frame-work has only two free parameters, $v$ and $H_0$,
 while standard cosmology
 actually uses three parameters, $\Omega_{m0}$, $H_0$ and $w$.
 The superiority of our frame-work over standard cosmology
 is mainly on theoretical aspects.

 Standard cosmology does not give any explanation of negative
 pressure's producing mechanism, so the equation of state coefficient $w$
 of dark energy must be counted as a free parameter.
 But we provide a possible negative pressure producing mechanism,
 it can produce $p=-\frac{1}{3}\rho$. We will discuss this
 problem in more details in the discussion section.
 Since we do not resort to dark energies to explain the
 luminosity-distance v.s. red-shift relation of super-novaes, our
 universe contains only matters today, so our cosmological
 frame-work has no coincidence problem \cite{Quintessence1}.

 Since in our cosmological frame-work, universe expands
 with zero acceleration, the size of the observable universe is
 always equal to the particle horizon of the universe. So our
 frame-work has no horizon problem.
 Since our cosmological frame-work has no horizon problem, quantum
 fluctuations inside the horizon will provide the primordial seeds
 for latter structure formations. So our frame-work has no
 primordial structure formation seeds problem \cite{Guth81,Linde82,EkpyroticUniverse}.

 Considering physics related with co-moving
 coordinate system, the global topology of the universe is
 not related with the energy density of the universe through
 a simple Friedmann equation
 $\frac{\dot{a}^2}{a^2}+\frac{k}{a^2}=\frac{8\pi G}{3}\rho_{tot}$.
 We have not found the metric of a closed/open universe by
 kinematical method. Probably, eq(\ref{metricDim3co}) is the only
 solution of the real universe. If that is the case, our
 cosmological frame-work has no flatness problem at all.

 \section{Discussions}
 There are two worries about our considering of physics related
 with co-moving coordinate system.
 The first is, since we use special relativity velocity addition
 rules to calculate recession velocity of galaxies on a given
 observational direction, some people worry that our theory will contradict
 the basic fact of Hubble's discovery, the recession velocity of
 a galaxy is proportional to the distance the galaxy being away from
 us, $v\propto H_0\cdot x_{ph}$. First let me explain that,
 even for this basic fact,
 different standard cosmologists could give us different interpretations.

 The first class standard cosmologists say that this is an empiric formulae
 only valid at low red-shift. Because if it is valid at very high
 red-shift, or on very large physical distances, super-light
 velocity of recessions would appear, which is anti-relativity.
 The second class standard cosmologists
 say that, $v\propto H_0\cdot x_{ph}$ is a basic principle
 valid on any scales; super-light recession velocity on super-horizon scales
 introduces no problem, so is allowed.
 We support the first class of standard cosmologists, i.e., $v\propto
 H_0\cdot x_{ph}$ is an empiric formulae only valid on low red-shift or
 small (compare with observational horizon of the universe)
 scales.

 Still take the one-dimensional galaxy line in
 FIG. \ref{LorentzContraction} as our experimental labs.
 Consider the recession velocity and the physical distance of a
 galaxy located at $(t,x)$ relative to us,
 $x$ is the natural coordinate of that galaxy,
 \beq{}
 v_x&&\hspace{-3mm}=\frac{e^{\sigma x}-e^{-\sigma x}}{e^{\sigma x}+e^{-\sigma x}};
 \label{recessionVelocity3}\\
 x_{ph}&&\hspace{-3mm}=\frac{4a}{\text{ln}\frac{1+v}{1-v}}\left[\text{arctg}[(\frac{1+v}{1-v})^{\frac{X}{2}}]-\frac{\pi}{4}\right],
 \label{physicalDistance3}
 \eeq
 Obviously, only when $v_x<<1$, i.e. $\sigma x<<1$, $v_x\propto
 x_{ph}$. Obviously, regardless how large is $x$, the recession velocity
 $v_x$ cannot be larger than $1$, the light velocity.

 The second worry about our cosmological picture is, since we
 assume that the average value of relative recession velocity
 between two nearest galaxies is time independent. The Hubble
 parameter is also time independent. This is just an illusion.
 Still take the one-dimensional galaxy line as our experimental
 labs. Obviously, since the distance between two nearest galaxies
 increase linearly with time, Hubble parameter, decreases as
 $t^{-1}$ as time passes by. This is the same as standard
 cosmology's matter/radiation dominated eras' Hubble parameter
 evolution rules. So if we use eq(\ref{metricDim3co}) to trace
 back the history of our universe, we will not get result
 inconsistent with Big Bang Nucleon-synthesis of standard cosmology.

 Our final discussion is about the negative pressure $p=-\frac{1}{3}\rho$'s producing
 mechanism. As the first step, let me ask if we have
 a one-dimensional infinitely long uniform galaxy line,
 and if the system is at rest initially, will it
 collapse at self-gravitations? Professor Ed. Witten once told me,
 Einstein contemplated similar questions.
 He considered a three-dimensional uniform lattice
 system, by poisson equation $\nabla^2\phi=\rho$
 (in general relativity, there are similar equations
 which will give us the same conclusions), the system has
 a solution $\phi=\frac{1}{2}\rho x^2$. So for any galaxy
 not on the $x=0$ plane, it will receive a force pointing to that
 plane. As a result the system will collapse to that plane.
 Of course, the system could also has solution like
 $\phi=\frac{1}{2}\rho (x-x_0)^2$, which means that the system
 should collapse to the $x=x_0$ plane. So Einstein concludes
 that an isotropic and homogeneous matter dominated
 universe cannot have static solution. This is
 why Einstein introduced cosmological constant into his basic
 equations to get static solutions, as early as before Hubble
 discovered that our universe is expanding.

 However, we wish to express a modest suspicion that,
 Einstein may ignore an important thing. In a infinitely long
 uniform galaxy line, inter-gravitations among different galaxies can
 produce negative pressures. Imagine that, there is a galaxy in
 the line containing less matters comparing with other galaxies.
 In this case, galaxies on the left hand side of this less
 weighted galaxy will collapse and move to the left,
 while galaxies on the right hand
 side of this less weighted galaxy will collapse and move to the
 right. So the less weighted galaxy will receive gravitations from
 both sides, which have intentions to split this
 galaxy into two parts. This intention can be understood as
 origins of negative pressure $p=-\frac{1}{3}\rho$.
 Then why is the equation of state coefficient $-\frac{1}{3}$?

 Before answering this question, let us first imagine that, if
 what we illustrated in FIG. \ref{LorentzContraction} is not a
 galaxy line, but an electron-line.
 Will the system expand at self-repulsion? According to the same
 poisson equation analysis of Einstein, the system should
 not expand at self-repulsion, but should collapse at
 self-repulsions!
 This is un-acceptable. So, maybe Einstein, and
 almost all standard cosmologists since Einstein, were cheated
 by their intuitions: an infinitely long uniform galaxy line will
 collapse at self-gravitations. They analyzed this problem by first ignoring
 pressures originated from the inter-gravitations(or static
 electronic repulsions) among the composite objects of the
 system then using the so-called dynamic equations
 (Einstein equation or Poisson equation) so get their conclusions.

 Our point of view is, to answer the question that, will an infinitely long
 uniform galaxy line collapse at self-gravitations, or an
 electron line expanding at self-repulsions? We should not
 use dynamic equations at the condition of ignoring
 pressures originated from inter-actions among different
 composite objects. Otherwise, we will get un-acceptable
 conclusions, e.g., an infinitely long uniform electron-line
 will collapse at self-repulsions.

 We think the reasonable conclusion should be, from
 symmetry analysis, any galaxy on the line receives gravitations
 from both sides. The two-side gravitations cancel each other,
 so any galaxy on the line will not run
 close to its neighbors, i.e., the system will not collapse or expanding at
 self-gravitations or self-repulsions.
 If we insist this analysis, then an initial expanding
 non-perturbed universe will keep expanding at the same speed for
 ever. For such a universe, we have used kinematic method and
 derived its metric in eq(\ref{metricDim3co}), by Einstein
 equation, the energy momentum tensor corresponding with metric
 has pressure $p=-\frac{1}{3}\rho$.

 We must claim that, when we say Einstein and his following
 standard cosmologists were cheated by their intuitions, we only
 want to express our modest suspicion. We have received many many
 criticisms and lampoons for our suspicion. But we think this
 suspicion is worth being kept in mind. After all, this
 suspicions has given us a possible explanation of the
 luminosity-distance v.s. red-shift relation of type Ia
 super-novae. We wish further exploration of this suspicion, for
 example, perturbing eq(\ref{metricDim3Nat}) or
 (\ref{metricDim3co}) and studying the structure formation or
 cosmic micro-wave background anisotropy problem then comparing
 with experimental observations such as SDSS \cite{Tegmark04}
 and WMAP \cite{WMAP03} will tell us
 whether our suspicion can be fact or not.

 \section{Conclusions}

 We derive the metric of an expanding universe with zero
 accelerations by pure kinematic method. By doing so we expatiate
 physics related with co-moving coordinate system in details. The
 most important discovery of our study is, in an expanding
 universe with zero accelerations, the red-shift of photons from
 distance galaxies is determined by the co-moving coordinate of
 the source galaxy instead of the scale factor ration $\frac{a(t_0)}{a(t)}$.
 Our discovery is consistent with the current observed
 super-novaes's luminosity-distance v.s. red-shift relations.

 We also discuss that, an expanding universe with zero
 accelerations has no horizon problem, (probably)no flatness
 problem, no primordial structure formation's seed problem. By
 Einstein equation, we find that to assure an expanding universe
 with zero accelerations, then energy momentum tensor of the
 underlying cosmological fluid must have $p=-\frac{1}{3}\rho$. We
 discuss that such a negative pressure can originate from the
 inter-gravitations among different composite objects of the
 universe --- the galaxies.

 \begin{center}
 {\bf Acknownedgement\\}
 \end{center}
 Originally, this paper appears as an answering letter to
 criticisms on our works \cite{CosmoSDSF}. When we finish the first
 version of that paper, we send it to professor E. Witten, G. 't
 Hooft, P. J. Steinhardt and other peoples for comments and
 criticisms. They read our paper and give comments
 seriously. We thank them very much
 for their comments or criticisms on our work in that paper.
 Their reactions encourage us very much.

 The current version of this paper include results of discussions
 with professor L. Liu, S.-y Pei. They inquire me to give a talk
 on the topic discussed in \cite{CosmoSDSF} at Beijing Normal
 University. When I finish the demonstrating document for the talk,
 I find the document itself may express my ideals more clearly
 than the original paper. So I decide to update the answering
 letter with the current paper --in fact-- the demonstrating
 document of talk to be given in BNU.

 Although I ask so many people to read my paper and give comments
 or criticisms on it, and they indeed do so. This does not mean that
 they agree with me on my opinions. So none of these people is
 to take response for the errors in the paper. But if there is any
 reasonable points in the paper, I must owe the credit to all of
 them.

\end{document}